\documentclass[preprint,showpacs,aps,preprintnumbers,amsmath,amssymb,
nofootinbib]{revtex4}
\usepackage{epsfig}
\begin{document}

\begin{flushright}
\end{flushright}


\newcommand{\be}{\begin{equation}}
\newcommand{\ee}{\end{equation}}
\newcommand{\bea}{\begin{eqnarray}}
\newcommand{\eea}{\end{eqnarray}}
\newcommand{\bers}{\begin{eqnarray*}}
\newcommand{\eers}{\end{eqnarray*}}
\newcommand{\nn}{\nonumber}
\newcommand\un{\cal{U}}
\def\dis{\displaystyle}
\def\B{B_d^0}
\def\Bb{\bar{B}_d^0}
\def \drho{\bar \rho}
\def \deta{\bar \eta}
\def\l{\lambda}

\title{\large Probing the unparticle signal in  $b \to d $ penguin
processes }
\author{R. Mohanta$^1$  and A. K. Giri$^2$  }
\affiliation{$^1$ School of Physics, University of Hyderabad,
Hyderabad - 500 046, India\\
$^2$ Department of Physics, Punjabi University,
Patiala - 147002, India}

\begin{abstract}
We investigate the effect of unparticles in the pure $ b \to d $
penguin processes $ B^0 \to K^0 \bar K^0$ and $B^{+,0} \to \phi
\pi^{+,0} $. Since these processes receive dominant contributions
due to the {\it top} quark in the loop, direct and mixing-induced CP
asymmetry parameters in these processes are expected to be
vanishingly small in
the standard model. We find that due to the unparticle effect
sizable nonzero  CP violation could  be possible in these channels.
\end{abstract}

\maketitle

The standard model (SM) of electroweak  interaction is very
successful in explaining the observed data so far (with the exception of
neutrinos), but still it is
believed that the SM  is not the complete theory but rather a low
energy manifestation of a higher theory, the form which is yet
unknown. There exist many beyond the SM scenarios with interesting
consequences which are being tested at present or are likely to be
tested in the future experiments. In this context, it is expected
that the SM predictions will be subjected to intense scrutiny in the
upcoming experiments at the LHC or ILC to decipher the existence of new
physics, if any. The study of B physics provides us an opportunity
to test the SM predictions and to look for physics beyond the SM.

Recently a very promising idea has been proposed by Georgi
\cite{georgi} which could in principle exist and might have been
undetected so far. The possible signatures of it may be found at the
upcoming experiments such as the LHC. This fascinating idea, called
unparticle physics, has already taken the center stage and is believed
to be one of the viable new physics scenarios.

In the context of conventional particle physics the scale invariance
is broken which is very well described by the Quantum Field Theory
with most of the particles  having definite mass. But there could be
a hidden theory which is scale invariant, with a non-trivial infrared
fixed point, whose fields are known as Banks-Zaks (BZ) fields
\cite{bz}. It is further assumed that the scale invariant theory
interacts very weakly with that of SM particles by the exchange of
very massive particles with the generic form $O_{SM}
O_{BZ}/M_{\un}^k$.  The renormalizable couplings of the $BZ$ fields
induce dimensional transmutation at some scale $\Lambda_{\un}$ where
the scale invariance appears. Below this scale the $ BZ$ operators
match onto corresponding unparticle operators leading to a new set
of interactions \bers
C_{\un}\frac{\Lambda_{\un}^{d_{\cal{BZ}}-d_{\un}}}{M_{\un}^k} O_{SM}
O_{\un}\;, \eers where $C_{\un}$ is a coefficient in the low energy
effective theory and $O_{\un}$ is the unparticle operator with
scaling dimension $d_{\un}$. Furthermore, $M_{\un}$ should be large
enough such that its coupling to the SM must be sufficiently weak,
consistent with the current experimental data. It is this feebleness
of the interaction for which the unparticle sector might have been
undetected so far.  Interestingly, the unparticle stuff with a scale
dimension $d_{\un}$ looks like non-integral number $d_{\un}$ of
invisible massless particles. The effect of unparticle stuff on low
energy phenomenology has been extensively explored in Ref.
\cite{ref1}.

A clean signal of the unparticle stuff can be inferred from various
analyses, e.g., the missing energy distribution in mono-photon
production via $ e^- e^+ \to \gamma \un $ at LEP2  and direct CP
violation in the pure leptonic $B^\pm \to  l^\pm \nu_l$ modes etc.
In this paper we would like to investigate the effect of unparticle
stuff in the  rare decay modes $ B^{\pm, 0} \to \phi \pi^{\pm,0}$
and $B^0 \to K^0 \bar K^0 $, which have only $b \to d$ penguin
contributions in the standard model. These processes are highly
suppressed in the SM, as they arise only at the loop level,
involving the CKM matrix element combinations $V_{qb} V_{qd}^*$ with
$q=u,c,t$, which are very small i.e.,  ${\cal O}(\lambda^3)$ in the
Wolfenstein parameterization, and thus provide an excellent testing
ground for new physics. Therefore, it is natural to expect that the
effect of unparticle stuff, if it exists, could show up, with
striking signals  in these channels.

In order to see the effect of unparticles in these channels let us
first briefly describe the various observables, which will be
measured in the upcoming LHCb experiments. The time dependent CP
asymmetry for $B^0 \to f $, where the final state $f$ stands for
$\phi \pi^0/K^0 \bar K^0$ which could be accessible from both $B^0$
and $\bar B^0$, can be given by \bea a_{CP}(t) &= & \frac{\Gamma
(\Bb(t) \to f)-\Gamma(\B(t) \to f)}{\Gamma (\Bb(t) \to
f)+\Gamma(\B(t) \to
f)}\nn\\
&=&S_{f} \sin \Delta m_{B_d} t-C_{f}\cos \Delta m_{B_d} t\;, \eea
where \be  S_{f}=\frac{2 {\rm Im(\lambda)}}{1+|\lambda|^2},
~~~~~C_{f}=\frac{1-|\lambda|^2}{1+|\lambda|^2}\;,\ee are the
mixing-induced and direct CP violating parameters. In the above expression
$\lambda$ corresponds to \be \lambda=\frac{q}{p}\frac{A(\Bb \to
f)}{A(\B \to f)}\;, \ee where $q$ and $p$ are the mixing
parameters, represented by the CKM elements in the SM as \be
\frac{q}{p}=\frac{V_{tb}^* V_{td}}{V_{tb} V_{td}^*}\sim e^{-2i
\beta}\;. \ee Since in the SM the decay mode $\B \to \phi \pi^0 (
K^0 \bar K^0) $ receives dominant contribution only from  $b \to d$
penguins with {\it top} quark in the loop,  one can generically
write the decay amplitude as \be A( \Bb \to f) = -\frac{G_F}{\sqrt
2}V_{tb}V_{td}^*~ P_t \;, \ee where $V_{ij}$ are the CKM matrix
elements which provide the weak phase information and $P_t$ is the
penguin amplitude arising from the matrix elements of the four quark
operators of the effective Hamiltonian. The amplitude for the
corresponding CP conjugate process is given as \be A( \B \to f) =
-\frac{G_F}{\sqrt 2}V_{tb}^*V_{td}~ P_t\;. \label{eq6} \ee So we
have
 \be
\lambda =\left (\frac{V_{tb}^* V_{td}}{V_{tb} V_{td}^*}\right )
\left (\frac{V_{tb} V_{td}^*}{V_{tb}^* V_{td}}\right )=1\;, \ee and
hence \be C_{f}=S_{f}=0\;. \ee Thus, if the measured CP violating
asymmetries in $B^0 \to \phi \pi (K^0 \bar K^0)$ deviate
significantly from zero then it would be a clear signal of new
physics. Moreover, the decay amplitude also receives some
contribution from the internal {\it up} and {\it charm} quarks in
the loop. Therefore, the CP violating parameters will not be
identically zero but will have small nonzero values. However,
including these contributions it is shown in Ref \cite{rm} that the
CP violating observables in the decay mode $B^0 \to K^0 \bar K^0$ is
rather small in the SM. Hence, in this analysis we will assume that
the standard model amplitude to be  dominated by the top quark
penguin.

Now, we will include the new contributions to the decay amplitudes
arising due to the unparticle stuff. It should be noted that,
depending on the nature of the original $ BZ$ operator $O_{ BZ}$ and
the transmutation, the resulting unparticle may have different
Lorentz structures. In our analysis, we consider only the vector
type unparticle exchange.  Under the scenario that the unparticle
stuff transforms as a singlet under the SM gauge group
\cite{georgi},  the unparticles can couple to different flavors of
quarks and induce flavor changing neutral current (FCNC) transitions
even at the tree level. Thus, the coupling of the vector-type
unparticles ($O_{\un}^\mu$) to quarks is given as \be
\frac{c_V^{q'q}}{\Lambda_{\un}^{d_{\un}-1}}\bar q'
\gamma_\mu(1-\gamma_5) q~ O_{\un}^\mu+h.c.\;, \label{cv} \ee where
$c_{V}^{q'q}$ are the dimensionless coefficients which in general
depend on different flavors. If both $q$ and $q'$ belong to up
(down) quark sector, FCNC transitions can be induced by  the above
effective interactions. We will consider these couplings
to be real so that the CP odd weak phase associated with the
unparticle couplings is zero. The propagator  for the vector unparticle is
given by \be \int d^4 x e^{i P \cdot x}\langle 0 | TO_{\un}^\mu(x)
O_{\un}^\nu (0)|0 \rangle = i \frac{A_{d_{\un}}}{2 \sin d_{\un} \pi}
\frac{-g^{\mu \nu} +P^\mu P^\nu/P^2} {(P^2)^{2-d_{\un}}}e^{-i
\phi_{\un}}\;. \ee where \be A_{d_{\un}}= \frac{16 \pi^{5/2}}{(2
\pi)^{2 d_{\un}}}
\frac{\Gamma(d_{\un}+1/2)}{\Gamma(d_{\un}-1)\Gamma(2d_{\un})}\;,
 ~~~~{\rm and} ~~~\phi_{\un}=(d_{\un}-2)\pi \;.\ee

After knowing the nature of interactions between the unparticles and
the quarks,  we are now interested to see how they will affect the
transition  amplitudes for the processes under consideration. Due to
the effect of vector like unparticle, the new contributions  to the
$B_d^0 \to \phi \pi (K^0 \bar K^0)$ decay amplitudes are given as
\be A(B^0 \to \phi \pi (K^0 \bar K^0))=- e^{-i
\phi_{\un}}~\frac{(c_V^{db} c_V^{ss})}{P^2} \frac{A_{d_{\un}}}{2
\sin d_{\un}\pi}\left (\frac{P^2}{\Lambda_{\un}^2} \right )^{
d_{\un}-1} X\;, \ee where
$X=\langle \phi \pi (K^0 \bar K^0)|(V-A)_\mu (V-A)^\mu|B^0
\rangle$ is the hadronic matrix element. In
the above equation we have taken the momentum transferred to the
unparticle as $P^2= m_{B_d} \bar \Lambda$ with $\bar \Lambda =
m_{B_d}-m_b$.

Now, including the unparticle contributions, one can write the total
amplitude as \bea A^T(\B \to f) = A^{SM}(1+r~ e^{i(\beta
-\phi_{\un})})\;, \eea where $A^{SM}$ is the SM amplitude as given
in (\ref{eq6}), $\beta$ is the weak phase associated with the CKM
elements $V_{tb}V_{td}^*$, $\phi_{\un}$ is the CP conserving strong
phase associated with the time-like unparticle propagator and $ r$
denotes the ratio of unparticle to SM amplitude, which is given as
\be r= \frac{c_V^{db} c_V^{ss}}{|V_{tb}^* V_{td}|}~\frac{ 1}{G_F
P^2}~\frac{A_{d_{\un}}}{ \sqrt 2 \sin (d_{\un} \pi)}\left
(\frac{P^2}{\Lambda_{\un}^2} \right )^{ d_{\un}-1} \frac{X}{P_t}\;.
\ee Thus, we obtain the CP averaged branching ratio $\langle {\rm
Br} \rangle \equiv [{\rm Br}(\B \to f)+{\rm Br}(\Bb \to  f)]/2$,
including the unparticle contributions, as \be \langle {\rm Br}
\rangle = {\rm Br}^{\rm SM} (1+r^2 +2 r \cos \beta \cos
\phi_{\un})\;, \label{eq15}\ee where $ {\rm Br}^{\rm SM} $
is the SM branching
ratio. The expressions for the CP asymmetries become \bea S_{f} &= &
-\frac{2 r \cos \phi_{\un} \sin \beta +r^2
 \sin 2 \beta}
{1+r^2+ 2 r \cos \phi_{\un} \cos \beta } \nn\\
\\
C_{f} &=&\frac{2 r \sin \phi_{\un} \sin \beta} {1+r^2+ 2 r \cos
\phi_{\un} \cos \beta }\;. \eea Thus, one can see that the branching
ratio and the CP violating  observables crucially depend on the
value of $r$ which in fact contains several unknown parameters i.e.,
the dimension of the unparticle fields $d_{\un}$, the energy scale
$\Lambda_{\un}$ and the couplings $c_{V}^{db}$, $c_{V}^{ss}$.
Therefore, it is not possible to constrain the new physics
contributions unless we fix some of these parameters. The coupling
constants $c_V^{db}$ can be constrained by the $B^0-\bar B^0$ mixing
data. Due to the unparticle exchange, the mass difference can be
explicitly given as \bea \Delta m_{B_d}= \frac{1}{2}\frac{f_{B_d}^2
\hat B_{B_d}}{ m_{B_d}} \frac{A_{d_{\un}}}{2| \sin d_{\un}\pi|}\left
(\frac{m_{B_d}}{\Lambda_{\un}} \right )^{2 d_{\un}-2}
|c_V^{db}|^2\;.\label{lk}
 \eea
Assuming that  the total contributions is given by the unparticles
and using  the result of $ \Delta m_{B_d}= 0.507~ {\rm ps}^{-1}$
\cite{pdg}, $f_{B_d} \sqrt{\hat B_{B_d}}$=0.2 GeV,
 the energy scale $\Lambda_{\un}$=1 TeV and the scale
dimension $d_{\un}$=3/2,
 one can obtain the upper bound on $c_{V}^{db}$ as \be
|c_V^{db}| \leq 2.3 \times 10^{-4}\;. \ee However, the variation of
the coupling $c_V^{db}$ with the scale dimension $d_{\un}$ for
$\Lambda_{\un}$=1 TeV, is shown in figure-1.
\begin{figure}[htb]
   \centerline{\epsfysize 2.0 truein \epsfbox{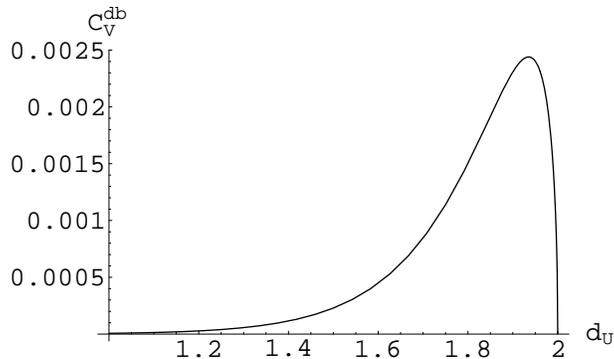}}
\caption{ Variation of $c_V^{db}$ with $d_{\un}$. }
  \end{figure}

Now let us first consider the decay mode $B^0 \to K^0 \bar K^0$. The
CP averaged branching ratio for this mode has already been measured
with value \cite{hfag} \be {\rm Br}(\B \to K^0 \bar K^0)=\left
(0.96_{-0.19}^{+0.21}\right ) \times 10^{-6}\;, \ee which agrees
with the SM predictions \cite{beneke}. The CP violating parameters
are  recently measured by both Babar \cite{bbr} and Belle \cite{bll}
collaborations and the world average values are \be \langle
S_{KK}\rangle=-0.82 \pm 0.55 ,~~~~~\langle C_{KK} \rangle =0.02 \pm
0.28. \ee
 Although, the measured branching ratio and the CP violation
 parameters (with large error bars) do not provide
any clear indication for a possible new physics effect, the precise
measurements of CP violation parameters in near future will
certainly establish/rule out the presence of new physics in this
channel.

The standard model amplitude for this process is given as \bea A(\Bb
\to K^0 \bar K^0) = -\frac{G_F}{\sqrt 2} V_{tb}V_{td}^*
\biggr[a_4-\frac{a_{10}}{2} +r_{\chi}\left (a_6-\frac{a_8}{2} \right
)\biggr]X\;,\label{smamp} \eea where $r_\chi=2
m_K^2/(m_b-m_s)(m_s+m_d) \approx 0.85$ is the chiral enhancement
factor and $X$ is the factorized matrix element given as \bea
 X = \langle  K^0 | \bar s
\gamma_\mu(1-\gamma_5)b | \Bb  \rangle
\langle \bar K^0 |\bar d
\gamma^\mu(1-\gamma_5)s|0 \rangle
 =  -i f_K F_0(m_K^2)~(m_B^2-m_K^2)
\;. \eea Now using the QCD coefficients $a_i$'s from \cite{lu}, the
value of the form factor $F_0(m_K^2)$  obtained using light cone QCD
sum rule approach \cite{ball}, the particle masses, lifetime of $B^0$ and $K$
meson decay constant $f_K=0.16$ GeV taken  from \cite{pdg}, the CKM
matrix elements as $|V_{tb}|=0.999125$, $|V_{td}|=8.72 \cdot 10^{-3}$
\cite{ckm}, we obtain the CP averaged branching ratio as
\be
{\rm Br}(B^0 \to K^0 \bar K^0)=8.6 \times 10^{-7}. \ee Although the predicted
branching ratio is in agreement with the experimental value, the
presence of new physics in this channel is not completely ruled out
unless the CP violating parameters are measured precisely, in
conformity with the SM expectations.

Now including the contributions arising from unparticle stuff we
show the variation of the CP averaged branching ratio
(\ref{eq15}), with the scale dimension
$d_{\un}$ in figure-2, where we have used the energy scale
$\Lambda_{\un}$=1 TeV, the value of $c_V^{db}$ for different $d_{\un}$
is extracted from the $B^0 - \bar B^0$ mixing data (as shown in Figure-1),
some representative set of values for $c_V^{ss}$ and
the weak phase  $\beta = 0.385$ rad. From the figure one can see that
the observed branching ratio
can be explained with unparticle physics for $d_{\un}> 1.2 $. As
$d_{\un}$ increases the branching ratio tends to the corresponding
SM value.
\begin{figure}[htb]
   \centerline{\epsfysize 2.0 truein \epsfbox{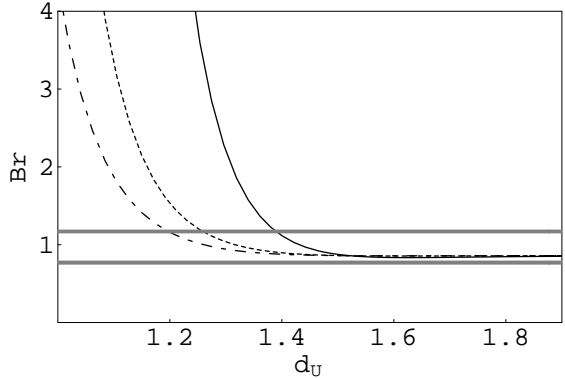}}
\caption{ CP averaged
  branching ratio $\langle {\rm Br} \rangle$ (in units of $10^{-6}) $ for
the decay mode $B^0 \to K^0 \bar K^0 $, where the solid, dashed and
dot-dased lines correspond to $c_V^{ss}$=0.05, 0.01 and 0.005
respectively. The horizontal thick lines represent the range of the
experimental data.}
  \end{figure}

The direct and mixing induced CP asymmetries are shown in Figure-3,
where it is found that  significant CP asymmetry could be possible
due to unparticle effect. As can be seen, large CP asymmetry is
possible for large  $c_V^{ss}$. And as $d_{\un}$ increases these
parameters tend to the corresponding SM values.
\begin{figure}[htb]
   \centerline{\epsfysize 2.0 truein \epsfbox{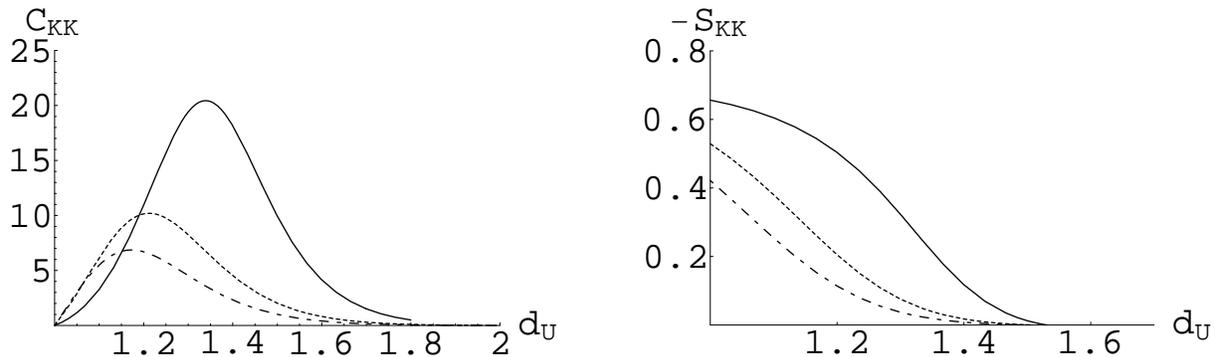}}
\caption{ Direct (in $\%$) and mixing induced
CP violation parameters for
the decay mode $B^0 \to K^0 \bar K^0 $, where the solid, dashed
and dot-dased lines correspond to $c_V^{ss}$=0.05, 0.01 and 0.005
respectively. }
  \end{figure}

Next we consider the processes $B^{+,0} \to \phi \pi^{+,0}$. These
modes have another interesting feature that they receive dominant
contribution from electroweak penguins as the strong penguins are
OZI suppressed. Hence they also provide an ideal testing ground to
look for NP. These modes have been analyzed in various beyond the
SM scenarios in Ref. \cite{rm1}. At present only the upper limits
of their branching
ratios are known \cite{hfag} \bea
{\rm Br}(B^+ \to \phi \pi^+) &<& 0.24 \times 10^{-6}\;, \nn\\
 {\rm Br}(B^0 \to \phi \pi^0) &<& 0.28 \times 10^{-6}\;.\label{eq24}
\eea

Let us first concentrate on $B^+ \to \phi \pi^+$ process. In the SM,
it receives contribution from the quark level transition $b \to d
\bar s s$, which is induced by the pure penguin diagram with
dominant contributions coming from electroweak penguins. Using the
generalized factorization approach one can write the transition
amplitude as \bea A^{SM}(B^+ \to \phi \pi^+) =- \frac{G_F}{\sqrt 2}
V_{tb}^*V_{td} \biggr[a_3+a_5-\frac{1}{2}\left ( a_7+a_9 \right )
\biggr]X\;,\label{eq:sm} \eea where \bea
 X &=& \langle  \pi^+ (p_\pi)| \bar d
\gamma_\mu(1-\gamma_5)b | B^+(p_B) \rangle
\langle \phi(q, \epsilon )|\bar s
\gamma^\mu(1-\gamma_5)s|0 \rangle \nn\\
& = & 2 F^{B \to \pi}_1(m_\phi^2)~f_\phi~ m_\phi~
(\epsilon \cdot p_B)\;.
\eea
is the factorized matrix element.
 The amplitude for
$B^0 \to \phi \pi^0$ is related to $B^+ \to \phi \pi^+$ by $A(B^0
\to \phi \pi^0)= A(B^+ \to \phi \pi^+)/\sqrt{2}$. The branching
ratio can be obtained using the formula \bea {\rm BR}(B^+ \to \phi
\pi^+) & = & \tau_{B^+} \frac{|p_{\rm cm}|^3}{
8 \pi m_\phi^2}~|{A(B^+ \to \phi \pi^+)}/({\epsilon \cdot p_B})|^2\;,\nn\\
{\rm BR}(B^0 \to \phi \pi^0) &=& \frac{\kappa}{2} ~ {\rm BR}(B^+ \to
\phi \pi^+)\;, \eea where $\kappa= \tau_{B^0}/\tau_{B^-}$ and
$p_{\rm cm}$ is the momentum of the outgoing particles in the $B$
meson rest frame.

For numerical evaluation we use the $\phi$ meson decay constant as
 $f_{\phi}=$ 0.237 GeV, the  form factor
 $F_1^{B \to \pi}(\phi^2)$ is obtained using QCD sum rule approach
\cite{ball} and the other parameters as presented for $B \to KK $
mode. Thus,  we obtain the branching ratio for $B^{+,0} \to \phi
\pi^{+,0}$ in the SM as \bea {\rm Br}(B^+ \to \phi \pi^+) &= &
3.95
\times 10^{-9}\;,\nn\\
{\rm Br}(B^0 \to \phi \pi^0) &=&1.85 \times 10^{-9}\;. \eea
These predicted values are quite below the present experimental
upper limits (\ref{eq24}).

Now including the unparticle contributions, the branching ratio and
the direct CP asymmetry parameters for the $B^+ \to \phi \pi^+$ are
plotted in Fig-4. From the figure one can see that the branching
ratio can be enhanced significantly and also large direct CP
violation could be possible in this channel. Since the direct CP
violation in the SM is identically zero, the observation of  nonzero
CP violation in this channel could be a direct signal of NP and the
unparticle stuff will be a strong contender of it.

\begin{figure}[htb]
   \centerline{\epsfysize 2.0 truein \epsfbox{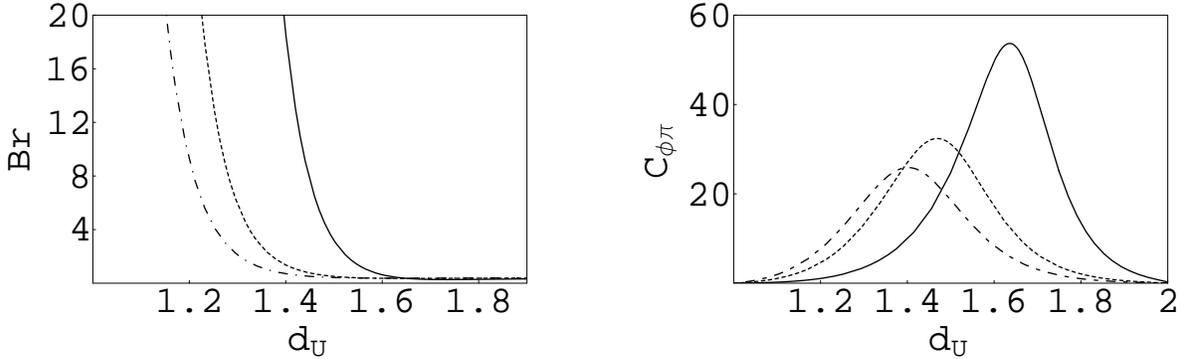}}
\caption{ CP averaged
  branching ratio $\langle {\rm Br} \rangle$ (in units of $10^{-8}) $ and
direct CP violation parameter (in $\%$) for
the decay mode $B^+ \to \phi \pi^+ $, where the solid, dashed
and dot-dased lines correspond to $c_V^{ss}$= 0.05, 0.01 and 0.005
respectively.}
  \end{figure}

One of the important goals of the B-factory is
to verify the standard model predictions and to serve as a potential
avenue to reveal new physics beyond the standard model (BSM).
Among the various BSM scenarios the recently advocated
unparticle physics scenario looks like a very strong
candidate indeed. In this context many interesting and novel
consequences have been pointed out in the literature which are likely
to be tested in the future experiments.
In this paper, we have explored some rare $b\to d$
penguin decay modes (namely, $B^0\to K^0 \bar K^0$ and $B\to \phi \pi$)
to study the effect of unparticle physics and possible
signatures of it. Specifically, in the case of
$B^0 \to K^0 \bar K^0$ mode although the branching ratio
appears to be in agreement with the
SM expectation but CP violating parameters can
reveal the existence of NP. It should be noted that the CP
violating parameters are close to zero in the
SM and nonzero values of the same, if found, will
clearly signal NP. Here we found that significant nonzero CP asymmetry
can be expected if unparticle effect
is taken into account. Similarly, in the case of $B\to
\phi \pi$ modes the branching ratios are very
small (only upper limits have been obtained so far)
and the contribution due to the unparticles can enhence the branching
ratios to significant ones. The direct CP violation in $B^+\to \phi\pi^+$ is
also zero in the SM but because of the unparticles we can expect large
direct CP violation in this case. To conclude, we have presented here some
rare decay modes where the SM predictions can be altered significantly by the
inclusion of unparticle effect which may be tested in the upcoming
experiments.

\acknowledgments We would like to thank J. D. Olsen for useful correspondence.
The work of RM was partly supported by Department
of Science and Technology, Government of India, through grant No.
SR/S2/HEP-04/2005. AKG would like to thank Council of Scientific and
Industrial Research, Government of India, for financial support.

\end{document}